\documentclass{article}
\usepackage{amsmath,graphicx,mlspconf}
\usepackage{xcolor}

\usepackage{amssymb}
\usepackage{booktabs}
\usepackage{color, colortbl}
\definecolor{Gray}{gray}{0.95}

\copyrightnotice{979-8-3503-2411-2/25/\$31.00 {\copyright}2025 IEEE}

\toappear{2025 IEEE International Workshop on Machine Learning for Signal Processing, Aug.\ 31-- Sep.\ 3, 2025, Istanbul, Turkey}

\title{Input Conditioned Layer Dropping in Speech Foundation Models}
\name{%
   Abdul Hannan$^{\: \star  ,\: \dagger}$%
   \qquad Daniele Falavigna$^{\dagger}$%
   \qquad Alessio Brutti$^{\dagger}$\thanks{This work was partially funded by the PNRR project CN - HPC (Spoke 9) under the NRRP MUR program funded by the Next Generation EU.}%
}
\address{%
   $^{\star}$ University of Trento, Italy \\%
   $^{\dagger}$ Fondazione Bruno Kessler, Italy%
}

\begin{document}
%\ninept
%\linespread{0.9}
\maketitle

\begin{abstract}
Curating foundation speech models for edge and IoT settings, where computational resources vary over time, requires dynamic architectures featuring adaptable reduction strategies. One emerging approach is layer dropping ($\mathcal{LD}$) which skips fraction of the layers of a backbone network during inference to reduce the computational load. This allows transforming static models into dynamic ones. However, existing approaches exhibit limitations either in the mode of selecting layers or by significantly modifying the neural architecture. To this end, we propose input-driven $\mathcal{LD}$ that employs the network's input features and a lightweight layer selecting network to determine the optimum combination of processing layers. Extensive experimentation on 4 speech and audio public benchmarks, using two different pre-trained foundation models, demonstrates the effectiveness of our approach, thoroughly outperforming random dropping and producing on-par (or better) results to early exit.
\end{abstract}
\begin{keywords}
dynamic models, layer drop, speech recognition, transformer
\end{keywords}
\vspace{-2mm}
\section{Introduction} \label{introduction} \vspace{-1.5mm}
Speech foundation models are extensively employed in diverse applications due to their efficient and rich semantic feature representation capability. However, their practicality on low resources / edge devices is limited due to significant computational overhead and enormous memory requirement. To address this issue, efforts have been directed to reduce the resource utilization of large models using different methods, such as pruning \cite{10.1145/3534678.3539260}%, peer2022greedy}
, quantization \cite{kim-hassan-2020-fastformers}, and low-rank decomposition \cite{winata2020lightweight, lirias3769084, kashiwagi23b_interspeech}. Although these approaches are fairly effective, they generate fixed architectures that fail to perform in varying resource settings. To this end, split-computing \cite{10096914, matsubara2022split} and early exit (EE) methods \cite{yoon24_interspeech, zaiem2023fine, wright2023training} have been introduced allowing the large models to adapt to varying computational resources. However, split-computing %methods 
may suffer from latency and privacy issues that degrade their effectiveness, while \textit{EE} methods require both careful selection of exit locations and additional overhead for training auxiliary output layers.

Alternatively, inspired from Stochastic depth approach \cite{huang2016deep}, \textit{Layer Dropping} ($\mathcal{LD}$), or Layer Skipping was proposed in \cite{Wang2018SkipNet, DBLP:conf/iclr/FanGJ20} as a sort of structured pruning technique that drops (or skips) complete layers or modules. Besides improving the training time~\cite{zhang2020accelerating}, random layer dropping (RD) has been employed in NLP with pretrained foundation models, i.e. BERT, RoBERTa, and XLNet architectures, by dropping pre-defined subnets~\cite{sajjad2023effect}. In the Automatic Speech Recognition (ASR) context, \cite{zaiem2023fine} utilized probabilistic $\mathcal{LD}$ to fine-tune WavLM \cite{chen2022wavlm} with a dropping probability of $0.5$. Following that, LDASR \cite{hannan2024} investigated the training and inference behavior of random (or probabilistic) $\mathcal{LD}$ on a conformer-based architecture using a range of dropping probabilities. These studies highlight that $\mathcal{LD}$ provides fair reduction in model size and also induces computational efficiency by skipping layers on random. We hypothesize that RD results in an ineffective combination of selected layers, producing a suboptimal model especially when the available resources are scarce (dropping most of the encoder layers). 
% Inspired from the self-attention mechanism that differently weighs parts of the input sample, we propose an \textit{input-driven} $\mathcal{LD}$ mechanism using a lightweight \textbf{Layer Selecting (LS)} block that identifies the optimal sub-network inside foundation model for each input sample and drops the least important layers. The LS block that operates in plug-and-play fashion, seamlessly integrates with various foundation models governing which model layers to skip or execute by assigning binary weights to the layers. Although data-driven approaches \cite{Wang2018SkipNet, ConvAIG, wright2023training, peng2023i3d} have been used in vision and NLP domains, either their applicability to transformer-based architectures for speech applications, especially ASR, is not investigated or the data-driven approach is unsuitable for extreme cases (dropping 60+\% of layers). Our approach transforms foundation models in to dynamic architectures as well as maintains efficient performance for varying resources even for extreme cases. In general, our contributions are as follows.
To this end, we propose an \textit{input-driven layer dropping} \textbf{(IDLD)} mechanism using a lightweight \textbf{Layer Selecting (LS)} block to transform foundation models in to dynamic architectures suitable for varying resources as well as to exhibit efficient performance even for extreme cases (dropping most of the encoder layers). %The proposed input-driven mechanism is inspired from the self-attention mechanism that weighs different parts of the input sequence. 
The LS block that operates in plug-and-play fashion, seamlessly integrates with various foundation models governing which model layers to skip or execute by assigning binary weights to the layers, hence selecting the optimal sub-network inside the foundation model. Although data-driven approaches \cite{wright2023training, Wang2018SkipNet, hannan2024, ConvAIG, peng2023i3d} have been used in vision and NLP domains, either their applicability to transformer-based architectures for audio applications, especially ASR, is not investigated or the data-driven approach is unsuitable for extreme cases. In general, our contributions are as follows.
%############### Figure ####################
\begin{figure*}[t!]
    \centering
    \includegraphics[width=0.80\linewidth]{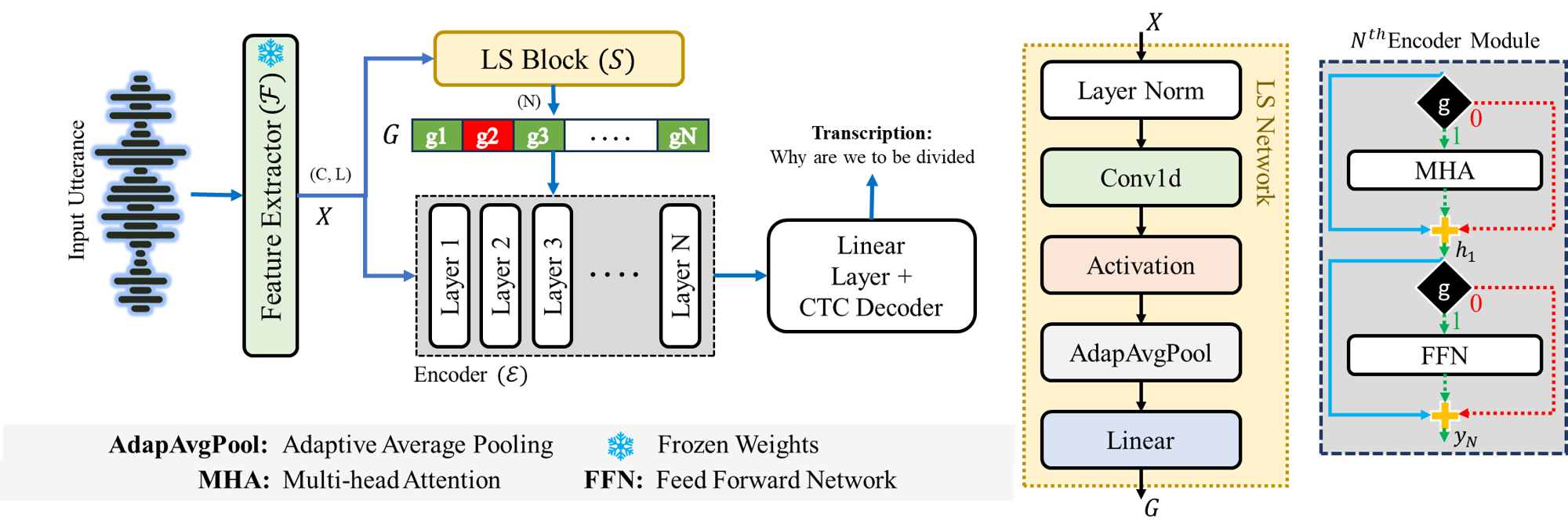}
    \vspace{-0.6mm}
    \caption{Illustration of the proposed input-driven layer skipping approach using a pretrained encoder. The output of the feature extractor $(X)$ is forwarded to both Encoder $\mathcal{E}$ and Layer Selector network $\mathcal{S}$. The latter provides layer scores $G \in \mathbb{R}^N$, for each layer of the encoder. Top-k values from $G=\{g^1, g^2, .., g^N\}$ are selected to enable corresponding encoder modules while skipping the remaining ones.}
    \label{fig:architecture}
    \vspace{-3mm}
\end{figure*}
%########################################

\begin{itemize}
    \item We propose input-driven layer dropping (IDLD) mechanism using a plug-and-play LS block, that enables optimal sub-network selection in foundation models for ASR and other audio applications.
    \item We demonstrate efficacy of the proposed mechanism for four downstream audio applications using two foundation models: \textbf{(i)} WavLM for ASR task on two public benchmarks, and \textbf{(ii)} Audio Spectrogram Transformer (AST) \cite{gong21b_interspeech} for sound classification, intent classification, and emotion recognition tasks. In addition, we compare threshold-based automatic layer selection with entropy-based early exiting.
    %\item We demonstrate efficacy of the proposed approach by exploiting WavLM model for ASR on two public benchmarks achieving superior performance as compared to the existing approaches. In addition, we evaluate our method on $3$ other audio applications i.e. sound classification, intent classification, and emotion recognition task, using Audio Spectrogram Transformer (AST) \cite{gong21b_interspeech}, highlighting its effectiveness.
    % \item We evaluate the foundation model to dynamically infer the dropping layers using a threshold-based mechanism that is analogous to entropy-based exiting in \textit{EE} approaches. 
\end{itemize}
\vspace{-3mm}
%\section{Background}
%\section{State of the Art} \label{sota}
\section{Background on Dynamic Depth} \label{sota} \vspace{-1.5mm}
Dynamic architectures have the ability to adapt the network's computational flow and can be categorized as: (i) dynamic depth \cite{Wang2018SkipNet, DBLP:conf/iclr/FanGJ20, zhang2020accelerating, sajjad2023effect, wright2023training, peng2023i3d, hannan2024}, (ii) dynamic width \cite{8578843, ConvAIG}, and (iii) dynamic routing \cite{fedus2022switch, jacobs1991adaptive}. We will restrict the discussion to dynamic depth only as it encapsulates the early exit and layer dropping approaches. Further details on other dynamic networks are available in~\cite{Han2022Survey}. The former offers multiple auxiliary exit branches to limit the processing beyond certain depth. Conversely, layer dropping skips intermediate layers (not necessarily contiguous) %of the network, 
adapting the network on the fly without extra classifiers. Next, we discuss the layer dropping approaches in more detail.
%\noindent \textbf{Dynamic Architectures }modify the computational flow of a neural network depending on the input sequence and can be categorized as: \textbf{(i) dynamic depth:} skipping parts of the backbone  \cite{Wang2018SkipNet, DBLP:conf/iclr/FanGJ20, zhang2020accelerating, sajjad2023effect, wright2023training, peng2023i3d, hannan2024}; \textbf{(ii) dynamic width: } refers to selective processing along channel dimension in each layer \cite{8578843, ConvAIG}; \textbf{(iii) dynamic routing: } involves computational path selection in parallel networks \cite{fedus2022switch, zhou2022mixture, jacobs1991adaptive}.% \textit{Dynamic depth} is a viable option as it can be easily integrated with pretrained transformer models without architectural alterations and provides reasonable model reduction. A popular approach is early exits that avoids the processing beyond certain depth by utilizing multiple decoders on each exit branch. Conversely, layer dropping skips intermediate layers (not necessarily contiguous) of the backbone, adapting the network on the fly without extra classifiers. Next, we discuss the $\mathcal{LD}$ approaches in more detail.

\noindent \textbf{Layer Dropping/Skipping,} initially proposed for ResNet in \cite{He2016ResNet}, and used in stochastic depth \cite{huang2016deep} and Transformers \cite{Waswani2017Transformer}, has been widely used in computer vision for efficient training. Examples of skipping strategies are: \textbf{(i) random dropping:} (used in NLP applications) where the blocks are dropped randomly while training sub-networks inside the backbone \cite{zaiem2023fine, DBLP:conf/iclr/FanGJ20, hannan2024, sajjad2023effect, Waswani2017Transformer}, \textbf{(ii) policy networks:} where a separate sub-network predicts the masking value for each block of the backbone network \cite{Wu2018BlockDrop, chen2019you, seon2023stop}, and \textbf{(iii) data-driven:} where the skipping decision is made on previous block's output \cite{Wang2018SkipNet, ConvAIG, Wang2020DualDynamic, Xia2020FullyDI}.

Recently, in vision domain, researchers put forward different approaches leveraging ResNet architecture to skip the model's layers. \cite{Wu2018BlockDrop} opted for reinforcement learning based skipping decisions whereas \cite{Wang2018SkipNet} used a hybrid of reinforcement and supervised learning to select the dropping layers on the basis of previous layer's output. \cite{seon2023stop} proposed to use a decision module for each layer separately, thus preventing to use effectively the available resources. Furthermore, \cite{chen2019you} suggested to process the input separately from the backbone network to compute which layers should be dropped/executed. Aforementioned, these methods are yet to be investigated for pre-trained transformer architectures without architectural alteration. For what concerns the ASR domain, work has been done using RD \cite{zaiem2023fine, hannan2024} and progressive layer skipping with a hybrid training methodology \cite{xu24_interspeech}, achieving reduced computational complexity coupled with fast convergence. In case of data-driven methods, \cite{wright2023training} employed EE, and \cite{peng2023i3d} employed $\mathcal{LD}$ with a different training recipe using Gumbel Softmax and skipping sub-modules of an encoder layer. Although effective, most of these studies do not work in plug-and-play fashion, the performance drop is significant for medium range dropping and they completely omit network's evaluation for extreme dropping cases.

\vspace{-2.5mm}
\section{Proposed Method} \label{sec:approach} \vspace{-1.5mm}
To mitigate the performance limitations without significantly altering the architecture of foundation model, we propose a lightweight, plug-and-play mechanism called as \textbf{input-driven layer dropping (IDLD)} using a \textbf{Layer Selecting} block as illustrated in Figure \ref{fig:architecture}. For each input sample, the LS block selects the finest combination of encoder layers achieving optimal performance for various resource settings.% To mitigate the performance limitations of Random Layer Dropping without significantly altering the architecture of foundation models, we propose a lightweight, plug-and-play mechanism called as  that selects the finest subnetwork from the foundation model for every input, hence, achieving optimal performance for various resource settings.

\noindent \textbf{Aim: } Given a dataset consisting of audio-label pairs $\{ (a^i, t_{ref}^i) \}_{i=1}^D$ with $D$ instances and a foundation model $\mathcal{M}$ consisting of a feature extractor $\mathcal{F}$, encoder $\mathcal{E}$ with $N$ layers, and a linear layer as a decoder, the objective is to transform the static model $\mathcal{M}$ in to a dynamic one, so that it can produce the best possible output irrespective of the number of encoder layers executed for the current sample.%To this end, we added a gating mechanism to the encoder $\mathcal{E}$ to enable skipping of $i$-th encoder layer. 

\noindent \textbf{Data Flow: }Figure \ref{fig:architecture} illustrates the overall framework of the proposed data-driven approach. The feature extractor $\mathcal{F}$ takes the $i^{th}$ audio sample $a^i$ to produce a feature representation $X^i = \mathcal{F}(a^i) $ that is utilized by encoder $\mathcal{E}$ and Layer Selecting block $\mathcal{S}$. The final logits $p^i$ are obtained by linearly projecting the output of $N^{th}$ encoder layer, and depending on the downstream application, either forwarded to a CTC decoder to get the transcription or is used to infer the target class.\vspace{-0.75mm}%The final embedding is obtained by linearly projecting the output of $N^{th}$ encoder layer, and depending on the downstream application, either forwarded to a CTC decoder to get the transcription or is used to infer the target class.\vspace{-0.75mm}
\begin{equation}
    p^i = \mathtt{Softmax} \big( f_\mathtt{linear} (\mathcal{E}(X^i)) \big)
    \label{eq:logits}
\end{equation}

\noindent \textbf{Static vs Dynamic Encoder: } For static models, each encoder layer consists of a multi-head attention (MHA) and feed-forward network (FFN) and can be formulated as:\vspace{-0.75mm}
\begin{equation} \label{eq1}
\begin{split}
    \hat{y}^j = y^{j-1} + f^j_\mathtt{MHA} (y^{j-1}) \\
    y^{j} = \hat{y}^j + f^j_\mathtt{FFN} (\hat{y}^j) \:\:\:\:\:
\end{split}
\end{equation}
where $y^{j-1}$ and $y^{j}$ are the input and output of encoder layer $\mathcal{E}^j$ ($j=\{1\dots N\})$. To allow skipping encoder layers, we add a gating mechanism (right most part of Figure \ref{fig:architecture}) that decides whether to skip or execute the complete encoder layer $\mathcal{E}^j$, instead of dropping separate sub-modules (MHA and FFN) as in~\cite{peng2023i3d}. So, equation \ref{eq1} can be written as:
\begin{equation}
\begin{split}
    \hat{y}^j = y^{j-1} + g^j \times f^j_\mathtt{MHA} (y^{j-1})  \\
    y^{j} = \hat{y}^j + g^j \times f^j_\mathtt{FFN} (\hat{y}^j) \:\:\:\:\:
\end{split}
\end{equation}
where the value of $g^j$ is given by Layer Selecting block. 

% %Consider a dataset $\mathcal{D}$ with the corresponding audio-label pair $\{ (a_d^i, t_d^i) \}$ at $i^{th}$ index containing $K$ total instances, we finetune a foundation model $\mathcal{M}_f$ to select the finest subnetwork for current input $a_d^i$ using proposed \textit{input-driven layer dropping (IDLD)} technique. Our method functions in plug-and-play fashion without modifying the architectural flow of the foundation model, which is unique from the past works in this domain.

% Figure \ref{fig:architecture} illustrates the overall framework using the proposed data-driven approach. For a dataset $\mathcal{D}$ with the corresponding audio-target pair $\{ (a_d^i, t_d^i) \}$ at $i^{th}$ index containing $K$ total instances, the feature extractor $\mathcal{F}$ of foundation model $\mathcal{M}_f$ processes the audio utterance $a_d^i$ producing an intermediate feature representation $X^i = \mathcal{F}(a_d^i) $ that is utilized by encoder $\mathcal{E}$ and Layer Selecting network $\mathcal{S}$. The encoder $\mathcal{E}$ consists of $N$ submodules where each submodule $\mathcal{E}^i$ is equipped with a gating mechanism (see right most part of Figure \ref{fig:architecture}) to decide whether to skip (red-arrow) or execute (green arrow). The final embedding is obtained by linearly projecting the encoder's output, and depending on the downstream application, either forwarded to a CTC decoder to get the transcription or is used to infer the target class. %employing a linear layer as a decoder at the encoder's output.

\begin{table*}[!t]
\centering
\caption{Dynamic behaviour depiction using WavLM model for (i) Automatic Speech Recognition, (ii) Intent Classification. (\textbf{n} - number of dropped layers, \textbf{RD} - Random Dropping, \textbf{IDLD} - Input-Driven Layer Dropping, \textbf{EE} - Early Exit)} %\vspace{-1.5mm}
\label{tab:wavlm}
\setlength{\tabcolsep}{9pt}
\begin{tabular}{@{}cccccccccc@{}}
\toprule
\rowcolor{Gray}
\textbf{} &
  \multicolumn{3}{c}{\textbf{LibriSpeech  (WER (\%))}} &
  \multicolumn{3}{c}{\textbf{TEDLIUM-v3}  (WER (\%))} &
  \multicolumn{3}{c}{\textbf{FSC}  (Accuracy (\%))} \\ \cmidrule(lr){2-4} \cmidrule(lr){5-7} \cmidrule(lr){8-10}
  \rowcolor{Gray}
\textbf{n} &
  \textbf{RD} &
  \textbf{IDLD} &
  \textbf{EE} &
  \textbf{RD} &
  \textbf{IDLD} &
  \textbf{EE} &
  \textbf{RD} &
  \textbf{IDLD} &
  \textbf{EE} \\ \midrule
0 &
  5.47 & 4.17 & \textbf{3.87} & 10.32 & 9.91 & \textbf{9.11} & 99.14$\pm$0.12 & \textbf{99.27$\pm$0.10} & 98.75$\pm$0.74\\
2 &
  5.78 & \textbf{4.22} & 4.63 & 10.75 & \textbf{9.96} & 10.66 & 99.19$\pm$0.13 & \textbf{99.57$\pm$0.07} & 98.69$\pm$0.73\\
4 &
  6.52 & \textbf{4.59} & 6.55 & 12.53 & \textbf{11.31} & 13.92 & 98.84$\pm$0.37 & \textbf{99.48$\pm$0.09} & 99.34$\pm$0.13\\
6 &
  9.01 & \textbf{6.08} & 9.09 & 17.74 & \textbf{15.09} & 18.20 & 98.63$\pm$0.26 & \textbf{99.24$\pm$0.08} & 99.33$\pm$0.20 \\
8 &
  17.59 & \textbf{11.24} & 14.84 & 31.90 & \textbf{26.33} & 27.06 & 94.76$\pm$0.37 & 95.23$\pm$0.07 & \textbf{98.98$\pm$0.23} \\
10 &
  60.10 & \textbf{38.75} & 38.85 & 76.76 & 63.52 & \textbf{55.91} & 60.59$\pm$0.88 & 74.61$\pm$0.44 & \textbf{97.64$\pm$0.16} \\ \bottomrule
\end{tabular}
\vspace{-4mm}
\end{table*}

\vspace{2mm}
\noindent The \textbf{Layer Selecting block }%: The Layer Selecting network, 
shown in Figure \ref{fig:architecture}, normalizes the feature representation $X^i$ of the sample $a^i$ and applies 1-D convolution with $C$ output channels followed by GELU activation. Afterwards, adaptive average pooling is applied to overcome input's dimensional variability (due to variable utterance length) followed by linear projection to get soft gate descriptors $\bar{g}^i \in \mathbb{R}^{N}$.% that are binarized using top-$k$ sampling to get $g^i$. The value of gate $g^i \in \{0, 1 \}$ governs whether to skip or execute the encoder $\mathcal{E}^i$.
\begin{equation}
%\begin{split}
    G = [\bar{g}^1,\dots,\bar{g}^N] = S(X) %\\
 %\end{split}
\end{equation} 

% \noindent During \textbf{training}, the value of $k$ in top-$k$ is sampled from uniform distribution $U(1, N)$ such that $1 \leq k \leq N$ allowing the model to be conditioned for various resource settings.

During \textbf{training}, the number of encoder layers per sample are selected using the top-$k$ values of $G$ where $k$ is sampled from uniform distribution $U(1, N)$. In this way, using top-$k$ method, the soft-gate values are transformed to binary gate descriptors $g^j$ of corresponding encoder layer $\mathcal{E}^j$.
\begin{equation}
       [g^1,\dots,g^N] = \text{top-}k \big( G, k\big) \: \text{where} \: k \in U(1, N) 
\end{equation}
% The $N^{th}$ encoder's output is linearly projected, and softmax is applied to get logits. 
% \begin{equation}
%     y_{logits} = Softmax \big( f_{linear} (y^N) \big)
% \end{equation}

For ASR, the logits $p^i$ of Eq.~\ref{eq:logits} are forwarded to a CTC decoder \cite{graves2006connectionist} to generate the transcription using $k$ encoder layers $t^i_{pred}|\mathcal{E}_{top-k}$, which is employed to estimate the CTC loss $\mathcal{L}_{CTC}$ to be minimized:
\begin{equation}
    \min \mathcal{L}_{CTC} \Big( t^i_{pred}|\mathcal{E}_{top-k}, \: t^i_{ref} \Big)
    % \mathcal{L}_{CTC} = \sum_{i}^{D} f_{CTC}\big( t_{pred}|\mathcal{E}_{topk}, \: t_{ref} \big)
\end{equation}
For other downstream tasks, the logits computed with $k$ encoder layers $p^i|\mathcal{E}_{top-k}$ are utilized directly to estimate the cross-entropy loss $\mathcal{L}_{CE}$:
\begin{equation}
    \min \mathcal{L}_{CE} \Big( p^i|\mathcal{E}_{top-k}, \: t^i_{ref} \Big)
 %    \mathcal{L}_{CE} = -\frac{1}{D} \Big( \sum_{i}^{D}\big[ (t^i_{ref})\:  
 % log(p^i|\mathcal{E}_{topk}) + (1-t^i_{ref})log(1- p^i|\mathcal{E}_{topk}) \big] \Big)
\end{equation}

\noindent At \textbf{inference}, we either set $k$ in top-$k$ to select a specified number of encoder layers or we define a threshold on the gate scores to select the relevant layers for a particular input (details in Sec \ref{sec:thres_selection}).

% \vspace{2mm}
% \noindent \textbf{Objective Function}
% $N^{th}$ encoder's output is processed and forwarded to ctc decoder to generate the transcription $t_{pred}|^k$, which is employed to estimate the CTC loss $\mathcal{L}_{CTC}$. The goal is to minimize $\mathcal{L}_{CTC}$ for each value of $k$ in top-$k$, so that $\mathcal{M}_f$ can produce the best possible transcription irrespective of the number of encoder layers executed for the current utterance.
% \begin{equation}
%     \min \mathcal{L}_{CTC} \big( y_{pred}|\mathcal{E}_{topk}, t_d \big)
% \end{equation}
% Without loss of generality, given an input $y^i$ to the $i$-th encoder block, the output of the block is given by:
% \begin{equation}
%     y^{i+1}={g}^i \cdot \mathcal{E}^{i}(y^{i})+y^{i}
% \end{equation}
% The $N^{th}$ encoder's output is processed and forwarded to ctc decoder to generate the transcription $t_{pred}|^k$, which is employed to estimate the CTC loss $\mathcal{L}_{CTC}$. The goal is to minimize $\mathcal{L}_{CTC}$ for each value of $k$ in top-$k$, so that $\mathcal{M}_f$ can produce the best possible transcription irrespective of the number of encoder layers executed for the current utterance.
% \begin{equation}
%     \min \mathcal{L}_{CTC} \big( y_{pred}|\mathcal{E}_{topk}, t_d \big) \: 1 \leq k \leq N
% \end{equation}

\section{Experimentation and Results}
\label{sec:experiments} 
We evaluated our proposed approach on $4$ downstream tasks 
\textbf{(i) Automatic Speech Recognition (ASR)} on LibriSpeech~\cite{panayotov2015librispeech} and TEDLIUM-V3~\cite{hernandez2018ted}, 
\textbf{(ii) Sound Classification (SC)} using Environmental Sound Classification 50 (ESC-50)~\cite{piczak2015dataset}, 
\textbf{(iii) Intent Classification (IC)} using Fluent Speech Commands (FSC)~\cite{DBLP:conf/interspeech/LugoschRITB19}, 
\textbf{(iv) Audio Emotion Recognition (ER)} using IEMOCAP~\cite{busso2008iemocap}. We utilized well-known transformer-based foundation models: (i) WavLM \cite{chen2022wavlm} for ASR and IC downstream tasks, (ii) Audio Spectrogram Transformer (AST) \cite{gong21b_interspeech} for SC, IC and ER. % downstream tasks. 
We employ Word Error Rate (WER) metric for ASR, and accuracy for the other tasks. We compare the IDLD approach against two baselines: RD with a dropping probability $0.5$ (gives best performance as shown in \cite{zaiem2023fine, hannan2024}) and data-driven EE approach. Our code is available online\footnote[1]{\textcolor{blue}{https://github.com/hannabdul/idld4fm}}.

\vspace{-1mm}
\subsection{WavLM} \label{wavlm-det-res} \vspace{-1mm}
\noindent \textbf{Implementation Details: }For \textbf{ASR}, we freeze the feature extractor of pretrained WavLM-base model and finetune the model for $50$ epochs. % using \textit{IDLD}, random dropping, and early-exits methods on LibriSpeech and TEDLIUM-v3 datasets. 
The training is performed on a NVIDIA $A40$ GPU with a batch size of 8 using Adam optimizer along with SpecAug~\cite{park19e_interspeech} with the default settings. We utilized the Connectionist Temporal Classification (CTC) loss for optimization while linearly increasing the learning rate followed by exponential decrease %till the end of training 
\cite{Waswani2017Transformer}. 
For \textbf{IC} task, we employed AdamW optimizer with high L2 regularization of $1e^{-2}$ to prevent overfitting on FSC dataset. We trained the model on $A40$ GPU for 20 epochs optimizing the cross entropy loss with a batch size of $32$ and initial learning rate of $1e^{-4}$ (decreased based on validation loss). The experiments are repeated $3$ times and average values are reported.

\noindent \textbf{Results:} Table \ref{tab:wavlm} enlists the obtained results in terms of: (i) WER for ASR on LibriSpeech test-clean and TEDLIUM-v3 corpus, and (ii) Accuracy for IC using FSC dataset. In case of \textbf{ASR} (left half of Table \ref{tab:wavlm}), IDLD outperforms RD for all values of dropped encoder layers $\text{n}$ (note that $n=N-k$), especially for $\text{n}=10$ and $8$ with a staggering difference of $27.66\%$ and $10.07\%$ on LibriSpeech. Similarly, IDLD outmatches EE for $\text{n} \geq 2$ showing the overall efficacy of input-conditioned layer dropping. %It should be noted that \textit{EE} method trains several auxiliary classifiers for each utterance whereas \textit{IDLD} has a solitary classifier working in conjunction with the LS Network at the top of network, achieving better performance. In addition, \textit{EE} optimizes the overall joint loss function comprising of all auxiliary classifiers that has been demonstrated to produce better results, whereas IDLD has a much simpler loss function conditioning the exclusive classifier to generate effective embeddings for all possible cases. Similarly, \textit{IDLD} has slight edge over \textit{EE} when using most of the encoder modules (like $n<2$) and out-matching EE for $n \geq 4$. 
In case of TEDLIUM-v3, we observe a similar trend as IDLD comfortably surpasses RD for all values of dropped encoder layers, with the difference becoming more evident for dropping $\text{n} \geq 6$ encoder layers.

For \textbf{IC} task, IDLD again surpasses RD in terms of performance for all values of n. The difference of $14.02\%$ in accuracy for $\text{n}=2$ using RD endorses our claim that input-conditioned dropping of layers yields significant improvement as compared to RD. However, when evaluated against EE it shows comparable (or better) performance for $\text{n} \leq 6$ while lagging behind for $\text{n}=8$ and $10$. 
%For \textbf{IC} task, \textit{IDLD} surpasses the \textit{random dropping} in terms of performance for all values of $n$. As discussed previously, the accuracy decreases by $14.02\%$ for $n=2$, endorsing our claim that conditioning dropping of encoder blocks to input utterance yields significant improvement as compared to random dropping. Furthermore, % \textcolor{red}{Comparison with EE coming soon} %Similarly, there is a staggering drop of $14.02\%$ in accuracy is observed when $n=2$ whereas the performance of both techniques is comparable for rest of the values of $n$. \textcolor{red}{compare with EE is missing}

%###############Figure ####################
\begin{figure}[t]
    \centering
    %\vspace{-5mm}
    {\includegraphics[width=0.75\columnwidth]{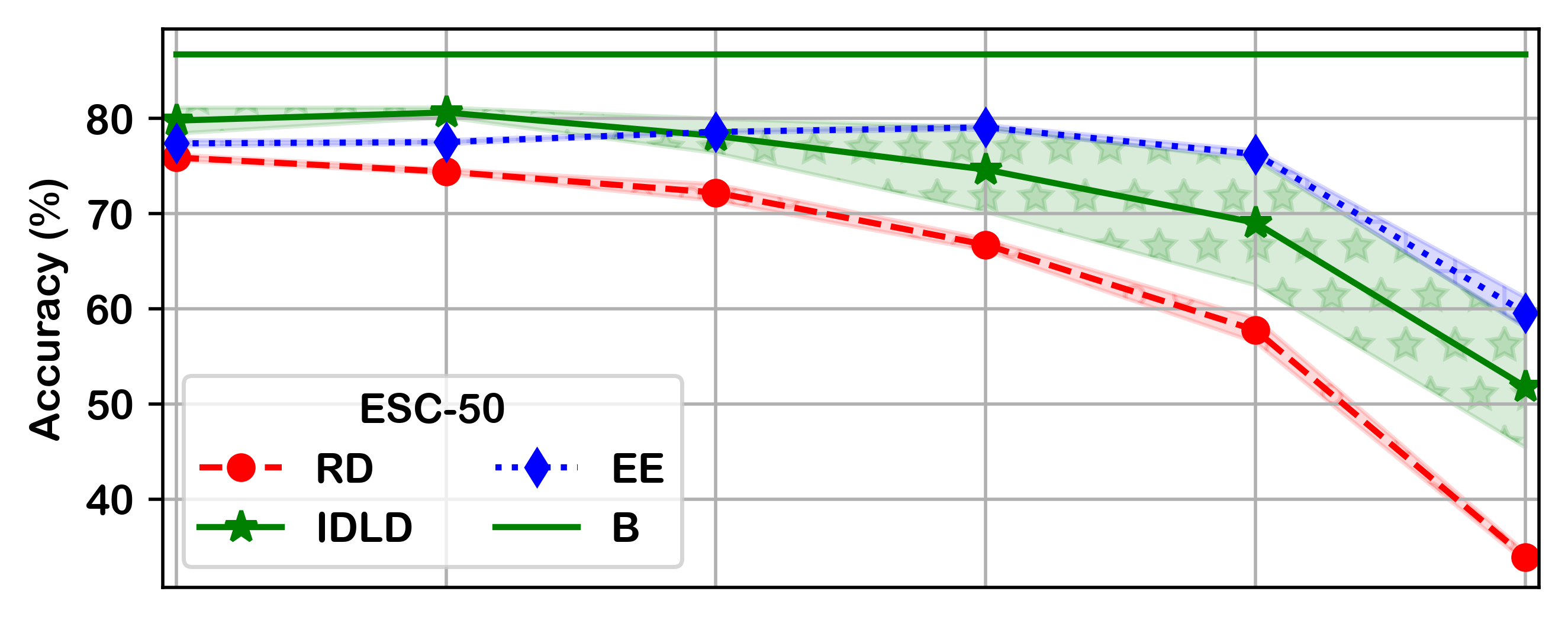}} %\vspace{-2mm}
    {\includegraphics[width=0.75\columnwidth]{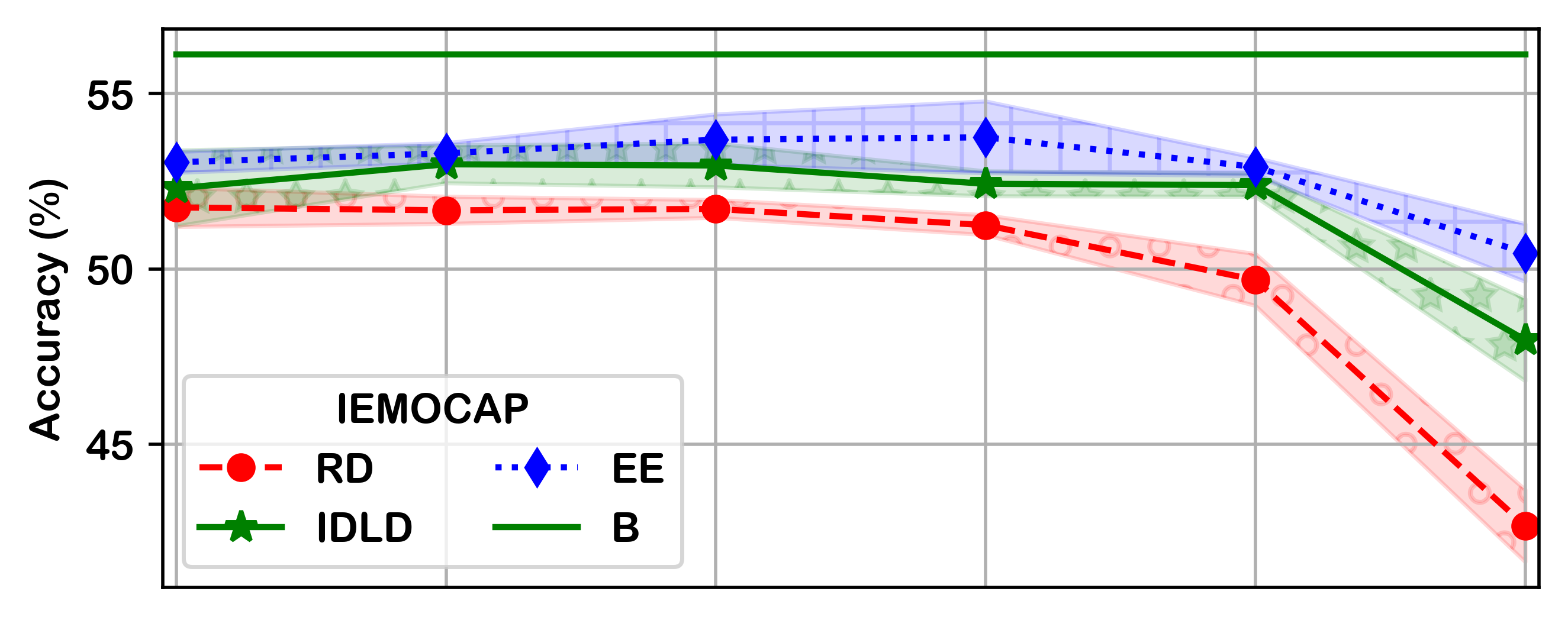}} 
    {\includegraphics[width=0.75\columnwidth]{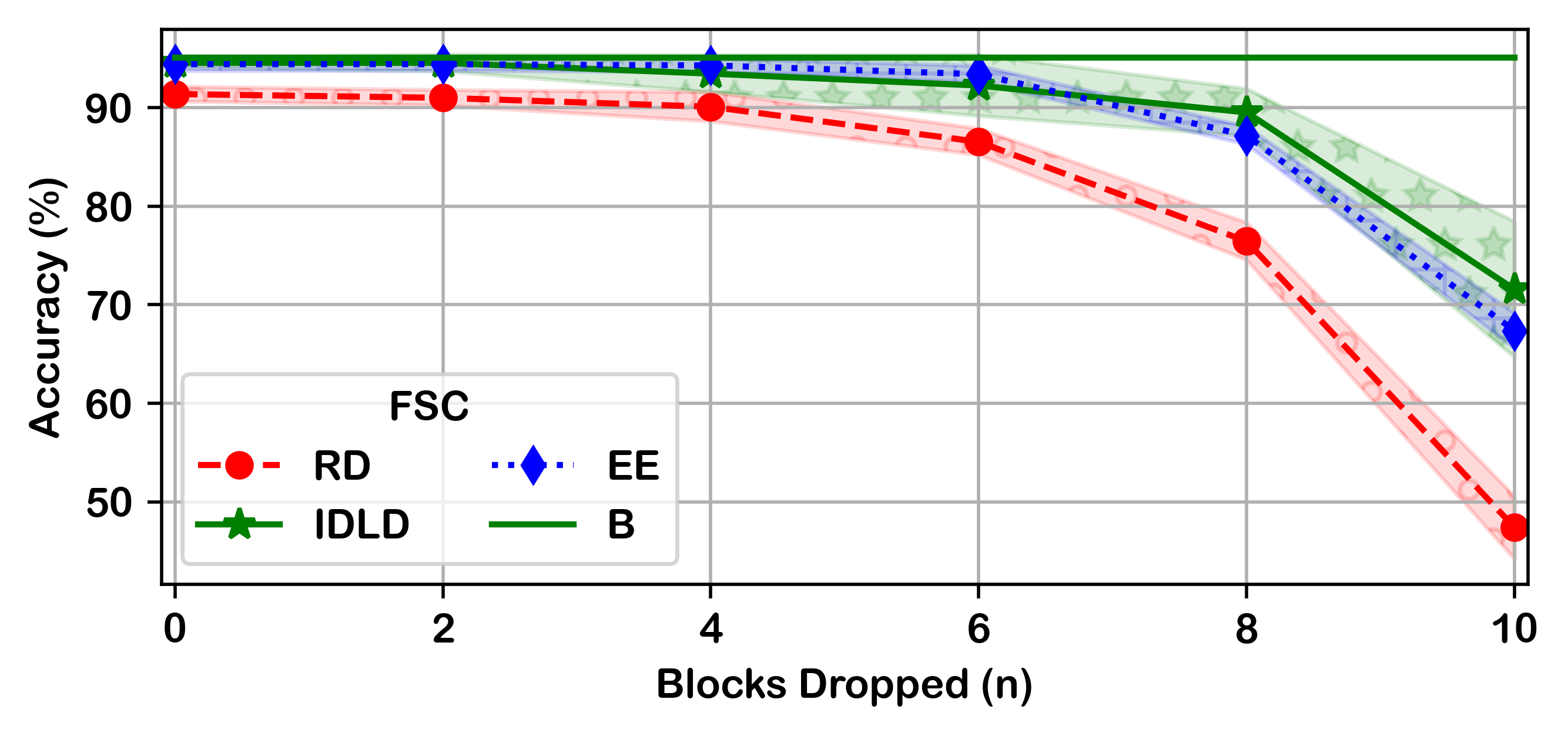}} \vspace{-1mm}
    \caption{Performance evaluation against the model's varying size using AST for (i) Sound Classification, (ii) Emotion Recognition, and (iii) Intent Classification. Light shaded color represents standard deviation of averaged runs.}
    \label{fig:ast_results}
    \vspace{-5mm}
\end{figure}

\subsection{Audio Spectrogram Transformer (AST)} \label{ast-det-res} \vspace{-1.5mm}
\noindent \textbf{Implementation Details: } For remaining tasks, we finetune the AST model pretrained on AudioSet using %Cross Entropy loss and 
AdamW optimizer while reporting averaged accuracy of $3$ runs. The training is performed on $A40$ GPU with $L2$ regularization of $1e^{-2}$ and initial learning rate of $1e^{-4}$ (decreased based on validation loss). As recommended in the respective articles, we applied 5-fold cross validation on ESC-50 dataset (\textbf{SC} task) and 10-fold cross validation on IEMOCAP dataset (\textbf{ER} task) for $50$ and $45$ epochs, while finetuning the model for $75$ epochs on FSC dataset (\textbf{IC} task). The training period (different number of epochs) lasted for different duration depending on different sizes of the datasets, and to allow the AST model to converge. %for random dropping, \textit{IDLD} and early exit methods. 
Furthermore, we empirically observed that the absence of a feature extractor is AST affects the extent of training required for model's convergence. Finally, to increase the available data for finetuning, we utilized SpecAug with frequency and time masking of 27 and 80 samples respectively.%we utilized SpecAug~\cite{park19e_interspeech} with a frequency masking of 27 samples and time masking of 80 samples.

\noindent \textbf{Results:} It is worth mentioning that AST being a purely attention-based classification model, doesn't contains a convolutional feature extractor which limits the learning of rich semantic representation required for LS block. Nevertheless, our method properly works with plain mel-spectrograms, and easily outperforms the random dropping as well as produce comparable results (even surpass in some cases) to early exit. Figure \ref{fig:ast_results} illustrates the observed variation in averaged accuracy using RD, IDLD, and EE methods. It is evident from Figure \ref{fig:ast_results} that input-driven layer dropping (solid green line) is far more superior than RD for all model sizes in each application, and the performance gap becomes more significant as n decreases ($\downarrow$ model size / $\uparrow \text{n}$). For instance, RD tries to keep up with IDLD for $\text{n} \leq 6$ but collapses after $\text{n} > 6$ due to the randomly selecting encoder modules. Quantitatively, RD exhibits $17.08\%$, $24.08\%$, and $5.31\%$ less accuracy than IDLD for $\text{n}=10$ on ESC-50, FSC, and IEMOCAP datasets. Conversely, when compared against EE, IDLD performs better for some values of n and lags behind for others on random across different applications. For example, for ESC-50 dataset, IDLD demonstrates a $5-$fold averaged accuracy of $80.61\%$ and $78.15\%$ as compared to $77.50\%$ and $78.57\%$ for EE, when dropping $2$ and $4$ encoder layers. The averaged accuracy for IDLD and EE closely follows each other with both surpassing one another in difference resource settings for ESC-50 and FSC datasets. Moreover, the minute difference in final accuracy for different number of dropped layers among IDLD and EE in different applications, highlights the technique's robustness with single classifier coupled with input-conditioning against EE with auxiliary classifiers.

\vspace{0.75mm}\noindent \textbf{Further Analysis of EE vs IDLD: }In Table \ref{tab:wavlm} and Figure \ref{fig:ast_results}, EE demonstrates good performance in all applications.  It should be noted that EE trains several specialized auxiliary classifiers and optimizes the overall joint loss function. Conversely, IDLD uses a solitary classifier with a simple loss function, working in conjunction with the LS block, to generate optimal output for all selected sub-networks. This fact favors EE when dealing with harder tasks or heavy mismatch between the pretrained model and the downstream task. Finally, note that due to the uniform sampling of $k$ during training, the IDLD model observes more often the lowest exits than the highest ones, being better optimized for them.

\vspace{0.75mm}\noindent \textbf{Performance drop for full model:} Figure~\ref{fig:ast_results} illustrates performance gap for proposed dynamic approach even when $\text{n} = 0$ with respect to an equivalent full static model (Baseline-B). This is an established fact in literature for dynamic models \cite{zaiem2023fine, wright2023training, hannan2024} as they are suitable for various resources.

\begin{figure}[t]
    \vspace{-0.5mm}
    \centering
    {\includegraphics[width=0.80\columnwidth]{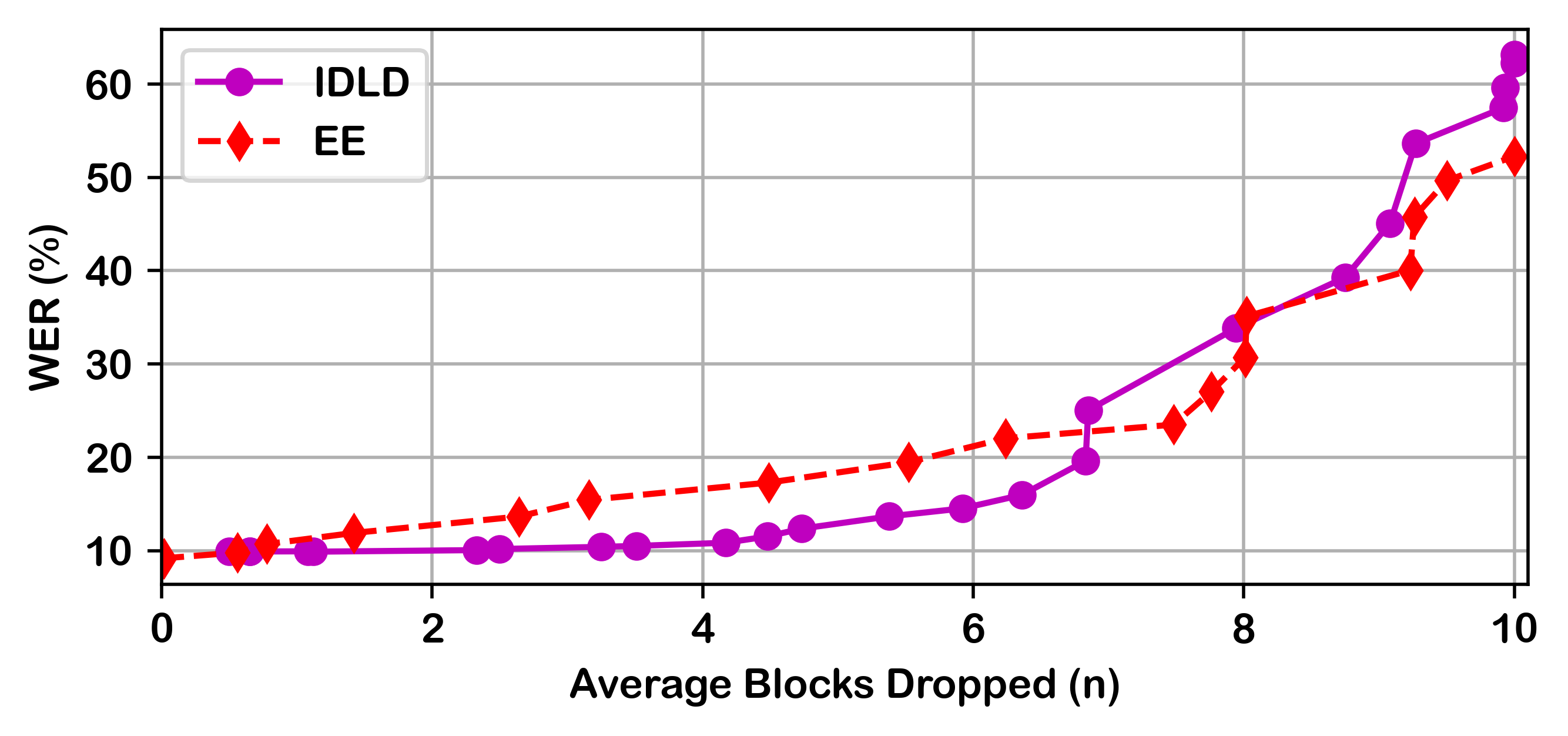}}
    \vspace{-4mm}
    \caption{Threshold based WavLM's encoder block selection on Tedlium test split.}
    \label{fig:thres}
    \vspace{-5mm}
\end{figure}

\noindent \textbf{Comparison with other works:} Note that \cite{peng2023i3d} implemented a similar layer skipping approach with a simple FFN layer selecting block and different training recipe, that is effective for a small dropping rate (up to 33.34\%) and shows a similar performance trend as in Figure \ref{fig:thres} on $10\%$ of LibriSpeech corpus. We experimented with a similar threshold-based recipe as in \cite{peng2023i3d}, and listed the results in Table \ref{tab:idld_rd} (Th-IDLD column) showing that top-$k$ approach is more favourable solution.%performs better than threshold-based approach.%showing the efficacy of our method.
%\vspace{0.75mm}\noindent \textbf{Performance drop for full model}. In Figure~\ref{fig:ast_results} there is a clear drop with respect to an equivalent full static model (Baseline-B) even when $\text{n} = 0$. This is an established fact in literature for dynamic models\cite{zaiem2023fine, wright2023training, hannan2024}.

% \vspace{-3mm}
% \subsection{Further Analysis of EE vs IDLD} \vspace{-1.5mm}

%EE -- specialized decoders for each case
%IDLD -- during training, n - 4-8 comes more often and it works better in these cases. 
%from other angle of which blocks will be dropped. Distribution of k.
\vspace{-3mm}
%############### Figure ####################
%##########################################
\vspace{-1mm}
\subsection{Threshold based selection} \label{sec:thres_selection} \vspace{-1.5mm}
At inference, besides using top-$k$ approach, the encoder layers required for each sample can be automatically inferred using a threshold-based mechanism. %, hence, removing the constraint on the number of encoder layer to select (as in top-$k$ approach).
To accomplish this, the soft gate values $\bar{g}^i \in \mathbb{R}$ are compared against a threshold ($\Gamma$) to deduce the best suited encoder layers for every sample, and binarizing $\bar{g}^i$ into $g^i \in \{0, 1\}$. Likewise, for EE model, the average entropy of the token's distributions is computed from each auxiliary exit, and the final output is taken from the lowest exit with entropy below the threshold value~\cite{wright2023training}. Figure \ref{fig:thres} illustrates the averaged encoder modules selected along with corresponding WERs when varying the gate-threshold value from $-0.6$ to $0.4$ for IDLD, and the entropy-threshold from $0.025$ to $0.7$ for EE. Note that similarly to the results in Table~\ref{tab:wavlm} and Figure~\ref{fig:ast_results}, IDLD is superior than EE in the middle and final exits while follows EE in the lowest exits. 

\vspace{-2.5mm}
\subsection{IDLD vs RD: Does LS Block really works ?} \vspace{-1.5mm}
To establish the superiority of IDLD over RD, we separately trained models with distinct dropping probabilities $p_{d-tr}$ that are suitable for different resources as well as variably selecting the dropping probability in the similar range as IDLD. Table \ref{tab:idld_rd} validates our claims that IDLD always perform better than random dropping, yielding better performance-computation trade-off for all model's sizes. In addition, a single model trained with our approach suffices individual models trained with random dropping for various resource settings, providing memory and performance efficiency.
%To establish the superiority of \textit{IDLD} over \textit{RD}, we separately trained models with distinct dropping probabilities $p_{d-tr}$ that are suitable for different resources as well as variably selecting the dropping probability in the similar range as \textit{IDLD}. Table \ref{tab:idld_rd} validates our claims that input conditioned dropping always perform better than random dropping, yielding better performance-computation trade-off for all model's sizes. In addition, a single model trained with our approach suffices individual models trained with random dropping for various resource settings, providing memory and performance efficiency.
%################################################
%\vspace{-2mm}
\begin{table}[t]
\centering
\caption{WER (in \%) for IDLD and Random  Dropping on LibriSpeech test-clean split using WavLM model} %\vspace{-1mm}
\label{tab:idld_rd}
\setlength{\tabcolsep}{4.75pt}
%\begin{tabular}{@{}lllllll@{}}
\begin{tabular}{ccccccc}
\toprule
\rowcolor{Gray}
\textbf{n} & \textbf{IDLD} & \textbf{Th-IDLD} & \multicolumn{4}{c}{\textbf{RD with $p_{d-tr}=$}} \\ \cmidrule{4-7}
\rowcolor{Gray}
 &  &  & \textbf{0.2} & \textbf{0.5} & \textbf{0.8} & \textbf{0.2 - 0.9} \\ \midrule
0 & \textbf{4.17} & 6.10 & 4.36 & 5.47 & 14.38 & 4.76 \\
2 & \textbf{4.22} & 5.17 & 5.12 & 5.78 & 13.78 & 5.13 \\
4 & \textbf{4.59} & 4.94 & 7.48 & 6.52 & 13.58 & 6.09 \\
6 & 6.08 & \textbf{5.83} & 14.79 & 9.01 & 14.62 & 8.36 \\
8 & \textbf{11.24} & 11.90 & 41.46 & 17.59 & 18.70 & 15.47 \\
10 & \textbf{38.75} & 62.82 & 94.05 & 60.10 & 39.45 & 45.91 \\ \bottomrule
\end{tabular}
\vspace{-5mm}
\end{table}

\section{Conclusion}
\label{sec:conclusions} \vspace{-2.5mm}
This work introduces a input-driven layer dropping (IDLD) approach for speech foundation models. The proposed approach utilizes a lightweight layer selecting block that selects the optimal sub-network depending on the model's intermediate representation per input, without altering the model's architecture. Moreover, it transforms the static foundation model in to a dynamic model that is scalable and functions in various resource settings with competent performance. Extensive experimentation using WavLM and AST foundation models on multiple audio applications, specifically ASR, using two benchmarks, reveal the merits of proposed approach in addition to superior performance-resources trade-off as compared to state-of-the-art methods.

\noindent \textbf{Limitations:} Table \ref{tab:wavlm} and Figure \ref{fig:ast_results} highlight the superiority of IDLD over RD in different audio applications. However, there is still room to further enhance the overall performance, primarily by tweaking the LS block's architecture. %incorporating recurrent neural networks or other modules in the LS network's architecture. 
Secondly, the output's standard deviation with IDLD method is greater than RD in AST model, hinting the reliance on deep feature extractor that is absent in AST. % model, and efforts are required to minimize the standard deviation achieving stable results while improving the robustness of LS module. 
Shortly, we believe that optimizing the LS block and minimizing the output's standard deviation will further boost our method's efficiency. % On top of that, even with better performance, it is observed that the standard deviation for IDLD method is greater than random dropping for AST model than WavLM model. There might be a link between the standard deviation of the output and the feature extractor of the used model. 

\bibliographystyle{IEEEbib}

\bibliography{strings,refs}

\begin{thebibliography}{10}

\bibitem{10.1145/3534678.3539260}
Sehoon Kim et~al.,
\newblock ``Learned token pruning for transformers,''
\newblock in {\em ACM}, 2022.

\bibitem{kim-hassan-2020-fastformers}
Young~Jin Kim et~al.,
\newblock ``{F}ast{F}ormers: Highly efficient transformer models for natural language understanding,''
\newblock in {\em Proceedings of SustaiNLP: WSENLP}, 2020.

\bibitem{winata2020lightweight}
Genta~Indra Winata et~al.,
\newblock ``Lightweight and efficient end-to-end speech recognition using low-rank transformer,''
\newblock in {\em ICASSP}. IEEE, 2020, pp. 6144--6148.

\bibitem{lirias3769084}
Pu~Wang et~al.,
\newblock ``Bottleneck low-rank transformers for low-resource spoken language understanding,'' 2022.

\bibitem{kashiwagi23b_interspeech}
Yosuke Kashiwagi et~al.,
\newblock ``{Tensor decomposition for minimization of E2E SLU model toward on-device processing},''
\newblock in {\em InterSpeech}, 2023, pp. 710--714.

\bibitem{10096914}
Arian Bakhtiarnia et~al.,
\newblock ``Dynamic split computing for efficient deep edge intelligence,''
\newblock in {\em ICASSP}, 2023.

\bibitem{matsubara2022split}
Matsubara et~al.,
\newblock ``Split computing and early exiting for deep learning applications: Survey and research challenges,''
\newblock {\em ACM Computing Surveys}, pp. 1--30, 2022.

\bibitem{yoon24_interspeech}
Ji~Won Yoon et~al.,
\newblock ``{H}u{BERT}-{EE}: Early exiting {H}u{BERT} for efficient speech recognition,''
\newblock in {\em Interspeech}, 2024.

\bibitem{zaiem2023fine}
Salah Zaiem et~al.,
\newblock ``Fine-tuning strategies for faster inference using speech self-supervised models: a comparative study,''
\newblock in {\em ICASSPW}. IEEE, 2023, pp. 1--5.

\bibitem{wright2023training}
George~August Wright et~al.,
\newblock ``Training dynamic models using early exits for automatic speech recognition on resource-constrained devices,''
\newblock {\em arXiv preprint arXiv:2309.09546}, 2023.

\bibitem{huang2016deep}
Gao Huang et~al.,
\newblock ``Deep networks with stochastic depth,''
\newblock in {\em ECCV}. Springer, 2016, pp. 646--661.

\bibitem{Wang2018SkipNet}
Xin Wang et~al.,
\newblock ``Skip{N}et: Learning dynamic routing in convolutional networks,''
\newblock in {\em ECCV}, 2018, p. 420–436.

\bibitem{DBLP:conf/iclr/FanGJ20}
Angela Fan et~al.,
\newblock ``Reducing transformer depth on demand with structured dropout,''
\newblock in {\em ICLR}, 2020.

\bibitem{zhang2020accelerating}
Minjia Zhang et~al.,
\newblock ``Accelerating training of transformer-based language models with progressive layer dropping,''
\newblock {\em NeurIPS}, pp. 14011--14023, 2020.

\bibitem{sajjad2023effect}
Hassan Sajjad et~al.,
\newblock ``On the effect of dropping layers of pre-trained transformer models,''
\newblock {\em Computer Speech \& Language}, vol. 77, pp. 101429, 2023.

\bibitem{chen2022wavlm}
Sanyuan Chen et~al.,
\newblock ``Wav{LM}: Large-scale self-supervised pre-training for full stack speech processing,''
\newblock {\em IEEE JSTSP}, pp. 1505--1518, 2022.

\bibitem{hannan2024}
Abdul Hannan et~al.,
\newblock ``{LDASR}: An experimental study on layer drop using conformer-based architecture,''
\newblock in {\em Proc. of EUSIPCO}, 2024.

\bibitem{ConvAIG}
Andreas Veit et~al.,
\newblock ``Convolutional {N}etworks with {A}daptive {I}nference {G}raphs,''
\newblock in {\em ECCV}, 2018, p. 3–18.

\bibitem{peng2023i3d}
Yifan Peng et~al.,
\newblock ``I3{D}: Transformer architectures with input-dependent dynamic depth for speech recognition,''
\newblock in {\em ICASSP}. IEEE, 2023, pp. 1--5.

\bibitem{gong21b_interspeech}
Yuan Gong et~al.,
\newblock ``{AST}: Audio spectrogram transformer,''
\newblock in {\em Interspeech}, 2021, pp. 571--575.

\bibitem{8578843}
Jie Hu et~al.,
\newblock ``Squeeze-and-{E}xcitation {N}etworks,''
\newblock in {\em CVPR}, 2018, pp. 7132--7141.

\bibitem{fedus2022switch}
William Fedus et~al.,
\newblock ``Switch transformers: Scaling to trillion parameter models with simple and efficient sparsity,''
\newblock {\em Journal of Machine Learning Research}, 2022.

\bibitem{jacobs1991adaptive}
Robert~A Jacobs et~al.,
\newblock ``Adaptive mixtures of local experts,''
\newblock {\em Neural computation}, pp. 79--87, 1991.

\bibitem{Han2022Survey}
Y.~Han et~al.,
\newblock ``Dynamic neural networks: A survey,''
\newblock {\em IEEE Transactions on Pattern Analysis \& Machine Intelligence}, vol. 44, no. 11, pp. 7436--7456, nov 2022.

\bibitem{He2016ResNet}
Kaiming He et~al.,
\newblock ``Deep residual learning for image recognition,''
\newblock in {\em CVPR}, 2016, pp. 770--778.

\bibitem{Waswani2017Transformer}
Ashish Vaswani et~al.,
\newblock ``Attention is all you need,''
\newblock in {\em NeurIPS}, 2017, vol.~30.

\bibitem{Wu2018BlockDrop}
Zuxuan Wu et~al.,
\newblock ``Block{D}rop: Dynamic inference paths in residual networks,''
\newblock in {\em CVPR}, 2018.

\bibitem{chen2019you}
Zhourong Chen et~al.,
\newblock ``You look twice: {G}aternet for dynamic filter selection in {cnn}s,''
\newblock in {\em CVPR}, 2019.

\bibitem{seon2023stop}
Jonghyeon Seon et~al.,
\newblock ``Stop or forward: Dynamic layer skipping for efficient action recognition,''
\newblock in {\em WACV}, 2023, pp. 3361--3370.

\bibitem{Wang2020DualDynamic}
Yue Wang et~al.,
\newblock ``Dual dynamic inference: Enabling more efficient, adaptive, and controllable deep inference,''
\newblock {\em IEEE JSTSP}, vol. 14, pp. 623--633, 2020.

\bibitem{Xia2020FullyDI}
Wenhan Xia et~al.,
\newblock ``Fully dynamic inference with deep neural networks,''
\newblock {\em IEEE Transactions on Emerging Topics in Computing}, vol. 10, pp. 962--972, 2020.

\bibitem{xu24_interspeech}
Jingjing Xu et~al.,
\newblock ``Dynamic encoder size based on data-driven layer-wise pruning for speech recognition,''
\newblock in {\em InterSpeech}, 2024, pp. 4563--4567.

\bibitem{graves2006connectionist}
Alex Graves et~al.,
\newblock ``Connectionist temporal classification: labelling unsegmented sequence data with recurrent neural networks,''
\newblock in {\em ICML}, 2006, pp. 369--376.

\bibitem{panayotov2015librispeech}
Panayotov et~al.,
\newblock ``Librispeech: an asr corpus based on public domain audio books,''
\newblock in {\em ICASSP}. IEEE, 2015.

\bibitem{hernandez2018ted}
Fran{\c{c}}ois Hernandez et~al.,
\newblock ``{TED-LIUM 3}: Twice as much data and corpus repartition for experiments on speaker adaptation,''
\newblock in {\em SPECOM}, 2018, pp. 198--208.

\bibitem{piczak2015dataset}
Karol~J. Piczak,
\newblock ``{ESC}: {Dataset} for {Environmental Sound Classification},''
\newblock in {\em Proceedings of the 23rd {Annual ACM Conference} on {Multimedia}}, 2015.

\bibitem{DBLP:conf/interspeech/LugoschRITB19}
Loren Lugosch et~al.,
\newblock ``Speech model pre-training for end-to-end spoken language understanding,''
\newblock in {\em Interspeech}. 2019, pp. 814--818, {ISCA}.

\bibitem{busso2008iemocap}
Carlos Busso et~al.,
\newblock ``{IEMOCAP}: Interactive emotional dyadic motion capture database,''
\newblock {\em Language resources and evaluation}, vol. 42, pp. 335--359, 2008.

\bibitem{park19e_interspeech}
Daniel~S. Park et~al.,
\newblock ``Spec{A}ugment: A simple data augmentation method for automatic speech recognition,''
\newblock in {\em Interspeech 2019}, 2019, pp. 2613--2617.

\end{thebibliography}

\end{document}